\documentclass[authoryear,final,3p,times,twocolumn]{elsarticle}

\usepackage{natbib}
\usepackage[]{hyperref}

\hypersetup{
    bookmarks=false,         
    unicode=false,          
    pdftoolbar=true,        
    pdfmenubar=true,        
    pdffitwindow=false,     
    pdfstartview={FitH},    
    colorlinks=true,       
    linkcolor=red,          
    citecolor=green,        
    filecolor=magenta,      
    urlcolor=cyan           
}

\usepackage{graphicx}

\usepackage{amssymb}

\usepackage{lineno}

\usepackage{xspace}
\usepackage{amsmath}  
\usepackage{textcomp} 

\journal{}

\begin{document}

\begin{frontmatter}



\title{{\textbf {Quantum mechanics needs no consciousness (and the other way around)}}}


\cortext[cor]{Current address: Laboratory of Systems Neuroscience, National Institute of Mental Health, 9000 Rockville Pike, Bethesda, MD 20892, USA}

\author[miph]{Shan Yu\corref{cor}}
\ead{yushan.mail@gmail.com}
\author[miph,fias]{Danko Nikoli\'{c}}
\ead{danko.nikolic@googlemail.com}

\address[miph]{Department of Neurophysiology, Max Planck Institute for Brain Research, Deutschordenstr. 46, 60528 Frankfurt am Main, Germany}
\address[fias]{Frankfurt Institute for Advanced Studies, Johann Wolfgang Goethe University, Ruth-Moufang-Str. 1, 60438 Frankfurt am Main, Germany}

\begin{abstract}

 It has been suggested that consciousness plays an important role in quantum mechanics as it is necessary for the collapse of wave function during the measurement. Furthermore, this idea has spawned a symmetrical proposal: a possibility that quantum mechanics explains the emergence of consciousness in the brain. Here we formulated several predictions that follow from this hypothetical relationship and that can be empirically tested. Some of the experimental results that are already available suggest falsification of the first hypothesis. Thus, the suggested link between human consciousness and collapse of wave function does not seem viable. We discuss the constraints implied by the existing evidence on the role that the human observer may play for quantum mechanics and the role that quantum mechanics may play in the observer\textquoteright s consciousness.
\end{abstract}





\end{frontmatter}


\section{Introduction}
\label{sec:intro}

The nature of human consciousness and its relation to the physical reality is arguably the most puzzling issue regarding the fundamental questions about ourselves and the interaction with the world that we live in. An interesting proposal has been put forward of a link between the seemingly distant quantum mechanics and consciousness, leading to a direct, yet bizarre bridge between the mental and the physical. It all started with the measurement problem in quantum mechanics, which can be formulated as follows:
According to quantum mechanics, the states of any physical system can be described fully by a wave function (state vector) that characterizes various system\textquoteright s variables such as its position, momentum, energy or spin. Schr\"{o}dinger\textquoteright s famous equation describes how these variables evolve over time \citep{Schrodinger_undulatory_1926}. According to most interpretations for the formalism of quantum mechanics (with the exception of the hidden variable theory, e.g., \citealp{bohm_suggested_1952}), the system described by the wave function does not have specific values (e.g., does not have a specific position), but is in a superposition state defined as the weighted sum of all states that the system may possibly assume following a measurement (known also as a set of eigenstates). This superposition can be verified experimentally, for example through interference phenomena \citep{zeilinger_experiment_1999}. However, for each single measurement, that is, whenever a macroscopic measuring device is used to detect the state of a particular system, the result always indicates a single eigenstate, e.g., a single photon always has a specific location in space. Importantly, the probabilities for observing the specific states, i.e. their distributions, are predicted most accurately by the wave functions, which describe the system as a superposition of multiple states prior to the measurement. This led physicists to conclude that a quantum system can evolve in two, very different, forms: one is continuous, deterministic and reversible, described by a wave function and occurs prior to the measurement. The other form is discontinuous but stochastic, as, during the measurement, the system \textquotedblleft jumps" suddenly from a superposition state into a single randomly chosen eigenstate. According to some interpretations of quantum mechanics, this jump is an irreversible event that occurs during the measurement process, and is usually referred to as the collapse of wave function or reduction of state vector. The measurement problem in quantum mechanics refers to understanding the nature of this \textquotedblleft collapse", both at the explanatory level, such as: \textquotedblleft Which other, more fundamental processes cause the collapse?", and the ontological level, such as: \textquotedblleft Is the collapse physically real or it is just an artifact of the theoretical system?". This measurement problem is a major topic of discussion in quantum physics and has been a source of disagreements among theoretical physicists for many years as there is a number of different ways in which one can interpret this set of theoretically very unsettling but empirically indisputable properties of quantum mechanics.

Von Neumann \citeyearpar{von_neumann_mathematical_1932} was probably the first person that addressed the problem of quantum measurement systematically and gave hints of its possible relation to human consciousness (for related reviews, see \citealp{primas_critical_1997,esfeld_wigners_1999,thaheld_does_2005,rosenblum_quantum_2008}). According to him, the measurement process consists of three main components: the system to be observed ($S$), the measuring instruments ($M$) and the observer ($O$). In order to measure the state of $S$, a physical device is needed. Let us denote it as $M_1$. This device has also its own states (e.g., the positions of the hand of a gauge) but these are sensitive to, and interact with $S$. The problem becomes more interesting when one realizes that, as $S$ and $M_1$ interact they form a combined system ($S^{\prime}$), which also needs to be observed. This observation can be made only by another measuring device ($M_2$), but then again $S^{\prime}$ in combination with $M_2$ forms $S^{\prime\prime}$, which needs yet another measuring device $M_3$, and the chain can go on up to the infinity. According to von Neumann, any measuring instrument, $M$, although a macroscopic object, should obey the fundamental rules of quantum mechanics, much like $S$. Thus, before the state of a device $M$ has been measured, the device must be in a superposition state, and this holds for every device in the chain, e.g., $M_1$, $M_2$, etc. This property postpones iteratively the collapse of the wave function to an ever later measuring device positioned higher and higher on the hierarchy, rendering thus the problem unsolvable. Von Neumann reasoned that in order to break this infinite chain of measurements and to give to the whole process a superposition-free, definite end, something with a very distinct property\textthreequartersemdash that cannot be described by the above procedure and hence, by the quantum mechanics\textthreequartersemdash needs to be involved. He also provided a formal proof that the formalism of quantum mechanics does not restrict the choice of the point at which such a \textquotedblleft cut" could be inserted and suggested, although only implicitly, that the \textquotedblleft subjective perception" of the human observer, $O$, or its
\textquotedblleft abstract ‘ego’" plays this important chain-braking role \citep{von_neumann_mathematical_1932}. Shortly after, London and Bauer \citeyearpar{london_la_1939} suggested explicitly that the collapse of the wave function and thus, the measurement of a quantum process, cannot occur without the registration of the results in the observer\textquoteright s consciousness. This new role of human consciousness in theoretical physics was defended by pointing out that consciousness has a \textquotedblleft completely special character", which is \textquotedblleft the faculty of introspection", and which in turn allows a person to be aware of the status of its own awareness \citep{london_la_1939}, corresponding to the measurement of itself and abrogating thus the need for any additional measurement devices. Therefore, the registration of a result in consciousness brings ultimately the initial system of the measurement into a new form\textthreequartersemdash taking a single eigenstate. Later, Wigner popularized this link between consciousness and collapse of wave function passionately \citep{esfeld_wigners_1999}. Wigner suggested that \textquotedblleft It is the entering of an impression into our consciousness which alters the wave function." and \textquotedblleft It is at this point that consciousness enters the theory unavoidably and unalterably."(cited from \citealp{shimony_role_1963}). Importantly, however, Wigner dropped this opinion completely at his final years \citep{esfeld_wigners_1999}.

Critical evaluation and heated debate on this hypothesis has not been absent (e.g., \citealp{putnam_commentspaper_1961, margenau_commentsprofessor_1962, shimony_role_1963, putnam_commentscommentscomments:reply_1964, cramer_transactional_1986, chalmers_conscious_1997, primas_critical_1997, mandel_quantum_1999, esfeld_wigners_1999, menskii_quantum_2000, brukner_youngs_2002, french_phenomenological_2002, thaheld_does_2005, koch_quantum_2006, penrose_road_2007, nauenberg_critique_2007, stapp_mindful_2007, rosenblum_quantum_2008}). Many of them address this issue from the philosophical point of view. Although they went to deep and interesting levels and brought up exciting ideas about fundamental aspects of the relationship between the mind and the physical world, those profound analyses failed to reach a simple and clear conclusion that would be widely accepted. Partly due to this reason, the hypothesis that consciousness causes (or is necessary for) the collapse of the wave function and, therefore, plays an important role in quantum mechanics remained a theoretical possibility for the interpretation of quantum mechanics. Although not preferred by most physicists, this solution to measurement problem is still strongly suggested even in some recent theoretical works (e.g., \citealp{menskii_quantum_2000, stapp_mindful_2007, rosenblum_quantum_2008}).

In the present paper, we do not aim to provide another philosophical argument. Instead, we attempt to address this issue from an empirical perspective. We re-formulate von Neumann\textquoteright s hypothesis as an empirically testable problem. We then attempt to falsify the hypothesis on the basis of the existing empirical evidence, as already suggested elsewhere \citep{mandel_quantum_1999, zeilinger_experiment_1999, brukner_youngs_2002}. In addition, we identify the experiments that need to be made in order to rule out alternative explanations and thus, to test the hypothesis more thoroughly.
This analysis is also informative for the study of consciousness itself, a phenomenon that is by no means easier to understand than the measurement problem \citep{chalmers_conscious_1997}. Following the hypothetical role of consciousness in the collapse of wave function, a \textquotedblleft symmetrical" proposal has been made, namely, that the collapse of wave function explains the emergence of consciousness. The most straightforward approach is to equate the consciousness with the collapse of the wave function. Therefore, through a bidirectional relationship, the two deep mysteries explain each other. This view has been explicitly expressed by e.g., \citet{mensky_reality_2007} but it is arguably an implicit assumption made by many other authors suggesting consciousness as a solution to the measurement problem (e.g. \citealp{von_neumann_mathematical_1932}; Wigner, as cited above; \citealp{london_la_1939, lockwood_many_1996}). Therefore, if the role of consciousness in the collapse of wave function can be falsified by empirical evidence, these suggestions of using quantum mechanics to explain consciousness will also become if not unwarranted, then considerably less attractive.

\section{Experimental design}
\label{sec:Experimental design}

First, let us formulate the hypothesis to be tested:
\vspace*{0.3cm}

{\it Proposition:} The event of forming an explicit phenomenal representation of
a result of quantum measurement in an individual observer\textquoteright s mind is necessary for the wave function (superposition state) of the system to collapse into a single eigenstate. 

\vspace*{0.3cm}

By using logic symbols of implication ($\Rightarrow$) we can write this statement formally as:

\begin{equation}
CWF \Rightarrow PR,
\label{eqn:CWF}
\end{equation} 

\noindent where CWF stands for \textquotedblleft collapse of wave function" and PR for \textquotedblleft phenomenal representation", meaning that the collapse of wave function should be always associated with a corresponding event of registering the results of measurement in consciousness. Or, by using logical negation ($\neg$) we can express this proposition equivalently as contraposition of (1):

\begin{equation}
\neg PR \Rightarrow \neg CWF, 
\label{eqn:CWF}
\end{equation} 

\noindent meaning that the collapse of wave function should never occur if the corresponding result of measurement has not been registered by a conscious observer.

There are multiple definitions of consciousness. Here we adopt a definition that can be operationalized. Therefore, the registration of a stimulus in the observer‘s consciousness means that the stimulus (i.e., the results of measurement) is perceived at the level of subjective experience and that the observer is aware of its presence such that he/she can produce an appropriate verbal report stating its identity. For example, one may state: \textquotedblleft The light beam hit the screen on the left side.", \textquotedblleft The oscilloscope showed 1 MHz signal." or \textquotedblleft The gauge pointed to 5 mV." We think that this definition is fundamentally consistent with the issues described in the introduction and is sufficient for the current analysis.

\begin{figure}
  \centering
      \includegraphics[width=0.45\textwidth]{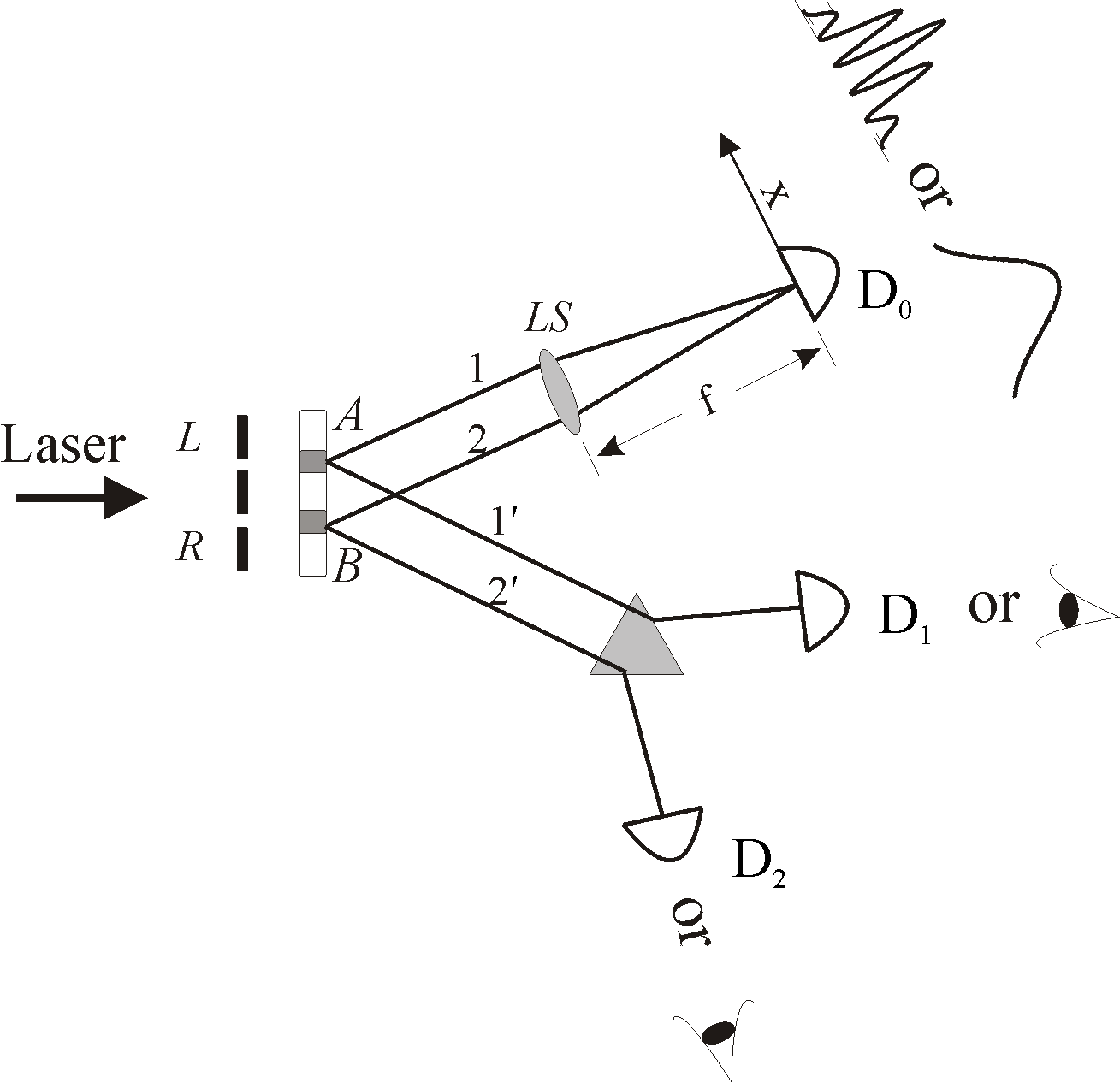}
     \caption{The proposed experimental setup that can be used to test whether collapse of wave function and consciousness about the outcome of the measurement are dissociated. This double-slit experiment is a modification of an actual experiment that has been carried out \citep{kim_delayed_2000} and is similar in principle to the setup proposed by \citet{scully_quantum_1982}. See main text for the detail. }
      \label{figure1}
\end{figure}

It is clear that this proposition can be never proven true, much like any other theoretical statement in science cannot ever be proven true \citep{popper_logic_1963}. However, propositions can be proven untrue, and in the present case this can be made simply by finding a counter example, that is, by finding an experimental setup in which the collapse of wave function is dissociated from consciousness about the outcome of the measurement. Thus, the first goal is to find an experimental setup that would allow one to assess both the state of consciousness, and, independently, the state of the wave function. To this end, we consider an adapted version \citep{kim_delayed_2000} of the experimental setup originally proposed by Scully and Dr\"{u}hl \citeyearpar{scully_quantum_1982} and designed to acquire \textquotedblleft which-path" information in a so-called double-slit experiment without interference (see Figure \ref{figure1}).

First, one photon from the pump travels through the double-slit and can hit either region A or B located on the nonlinear optical crystal to produce an entangled pair of photons. In the resulting pair, one photon (the signal) travels through the lens LS and is detected by the detector, $D_0$, positioned at the focal plane of LS. The other photon (the idler) is routed the other direction and travels through a prism to be diverted\textthreequartersemdash depending on the region in which it has been produced (A or B)\textthreequartersemdash either towards $D_1$ or $D_2$. Thus, by knowing which of the two detectors has registered a photon, we know which path the signal photon has taken.

Next, we analyze the system more closely. Assume first that the laser emits only one photon at a time. The state of the photon, $\Psi$ , can be described as:

\begin{equation}
\Psi =\frac{1}{\sqrt{2}} \big( |L \rangle + |R \rangle \big), 
\label{eqn:CWF}
\end{equation} 

where $|L \rangle$ and $|R \rangle$ indicate the photon's states, i.e., whether photon passed through the left or right slit, respectively. As a result, after the generation of a photon pair in the optical crystal, the signal photon may take either both paths 1 and 2 simultaneously (if it is in a superposition state, $|1,1' \rangle + |2,2' \rangle$ , and hence the wave function did not collapse) or through only one of the two (if it is in single state, $|1,1' \rangle$ or $|2,2' \rangle$ due to a collapse of the wave function). If the photons are always in a superposition state, after sufficient number of photons have been registered at $D_0$, the distribution of the registering location along the x-direction will exhibit standard Young's double-slit interference pattern, manifested by the distribution consisting of a series of peaks and troughs \citep{kim_delayed_2000}. In contrast, the photons that assume a single state will not produce such an interference pattern but will instead form a single-peak distribution \citep{kim_delayed_2000}. Thus, the presence of the interference pattern at $D_0$ indicates whether the wave function of signal photon collapsed or not \footnotemark. Thus, regarding the collapse discussed here, the relevant information (corresponding results of measurement) is which path the photons took. Now, we can derive the predictions for this experimental setup that follow form previously formulated {\it Proposition}:

\footnotetext {Interference pattern is not necessarily obvious from the entire distribution. So proper separation of sub-populations of registered photons may be needed \citep{kim_delayed_2000}.}

\vspace*{0.3cm}
The interference pattern should be visible if \textquotedblleft which-path" information has not been registered in consciousness of the observer (e.g., the experimenter).
\vspace*{0.3cm}

If the above is true, we expected to find the interference pattern at $D_0$ in the following conditions:

1) No actual attempt to measure the \textquotedblleft which-path" information was made, that is, $D_1$ and $D_2$ are not implemented at all.

2) The \textquotedblleft which-path" information was measured as $D_1$ and $D_2$ are implemented in order to interact with the incoming photons. However, no results were recorded by any macroscopic device and are not visible or in any way accessible to a human observer.

3) The \textquotedblleft which-path" information was measured by a macroscopic device such as $D_1$ and $D_2$. The results were not recorded but were instead presented to a human observer directly such that the relevant information entered the sensory system but, at the same time, the observer was distracted in order to prevent conscious detection of this event. In other words, the information necessary to achieve phenomenal representation was available in the nervous system, but conscious phenomenal experience was actually not realized. Thus, there were only non-phenomenal mental representations. The relevant information can be presented by using a memory-less device (e.g., an old fashion gauge-based instrument) or by feeding the idle photons (or after amplification) directly into the retina and, thus, having human eyes serve directly the function of $D_1$ and $D_2$ (see \citealp{brunner_possible_2008,thaheld_modified_2008,thaheld_mechanical_2009}). A distraction that prevents one from consciously detecting a stimulus is made routinely in psychological studies and can be achieved by various means, such as the visual masking \citep{lachter_disappearing_2000}, attentional blink \citep{raymond_temporary_1992}, binocular rivalry \citep{koch_quantum_2006}, change blindness \citep{rensink_to_1997}, execution of a concurrent tasks \citep{kahneman_perceptual_1967} , or by simply cluttering the visual scene \citep{treisman_feature-integration_1980}. The interference with conscious perception can be made even more directly by using trans-cranial magnetic stimulation. One could apply a magnetic pulse above e.g., visual cortex, in order to interrupt the information processing in this brain region, preventing hence conscious perception of the visual stimuli \citep{ silvanto_striate_2005}.

Moreover, one can manipulate gradually the level of subjective certainty of the presence of this information. This should then, according to the proposition, affect the contrast of the interference pattern accordingly\textthreequartersemdash as it has been shown in physical experiments by manipulating the extent to which the \textquotedblleft which-path" information was available, e.g., by changing the position of photon detector \citep{zeilinger_experiment_1999} or by attenuating the optical transmission \citep{mandel_quantum_1999}.

Verifying these three predictions through empirical tests we propose to be a necessary requirement to warrant the hypothesis that consciousness of the outcome of a measurement is necessary for the wave function to collapse. By formulating these predictions and requirement, we make this hypothesis empirically testable and hence, falsifiable.

\section{Existing evidence}
\label{sec:Existing evidence}

The experimental results necessary to falsify the predictions 1 and 2 already exist. First, as described by \citet{mandel_quantum_1999} and \citet{zeilinger_experiment_1999}, in experiments similar to that proposed here, if \textquotedblleft which-path" information was in principle obtainable, then even though no actual attempt was made to extract this information (i.e., to measure it), no interference pattern was found. Thus, the first prediction of consciousness hypothesis is false. In other set of experiments \citep{ eichmann_youngs_1993, durr_origin_1998}, \textquotedblleft which-path"
information was measured but was not recorded by any macroscopic device (for example, this information was stored only in the state of single atom or photon) and, therefore, was not accessible to a conscious observer. Under such condition, also no interference pattern was found. Therefore, the existing evidence indicates that the second prediction is also false.

To the best of our knowledge, no direct attempt was made to test the third prediction. However, the expectations for this experiment are clearly set by the evidence related to predictions 1 and 2. That is, if no interference pattern was obtained when the \textquotedblleft which-path" information was not fed into the eye of the observer (e.g., carried by the idler photon as illustrated in Fig.1), the same is expected to occur if the photon reached the observer\textquoteright s retina but the person was distracted as not to be able to detect the event.

\section{Discussion}
\label{sec:Discussion}

We first derived a proposition about the relationship between the collapse of the wave function and conscious perception. Our subsequent analysis lead to the conclusion that this proposition is already disproved by the existing empirical results, which forces us to conclude tentatively the following: Conscious access to the information about the outcome of a measurement of a quantum state is not necessary for the collapse of wave function\textthreequartersemdash conclusion similar to those suggested elsewhere \citep{mandel_quantum_1999,zeilinger_experiment_1999, brukner_youngs_2002}.

Does the present analysis really tell us something about the relation between consciousness and quantum mechanics? One may argue that with the current experimental set up (shown in Figure 1), quantum mechanics as we known can predict, correctly, no interference pattern in $D_0$, irrespective of what happen with the idler photons (except for \textquotedblleft erasing\textquotedblright \ the \textquotedblleft which-path\textquotedblright \  information that is carried by those photons, see \citealp {kim_delayed_2000}), and certainly irrespective of whether a conscious observer is involved and where the attention of this observer is directed. Therefore, this setup cannot tell us anything new about the relation between consciousness and quantum mechanics that we did not know before. So, did we just provide a circular argument?  

To answer this question, let us consider the type of experiments that can be proposed in principle. According to the opinion mentioned above, a really interesting test would involve some observables not determined by the current quantum theory. Only in that case, consciousness of the observer would be given a chance to affect the results and only in that case we would be searching empirically for novel discoveries. But, given the known properties of quantum mechanics, is it possible ever to conceptualize such an experiment?  Designing such an experiment would mean finding a situation in which the quantum mechanics is either incomplete (e.g., the current theory does not predict whether interference will be observed) or inconsistent (e.g., theoretically, presence or absence of interference are both possible). 

In the early years of quantum mechanics (see for example \citealp{einstein_can_1935}) doubts have been raised about the correctness of this theory\textthreequartersemdash which was a natural component of the scientific process. But by now, more than seven decades later, quantum theory has been proven to be one of the most accurate theories in the whole science. Not a single prediction of quantum mechanics has been empirically disproved. This casts doubts on the possibility of designing a novel experiment in which an observable is not completely constrained by the known theory and would be still open to the influences from the side of the consciousness of the observer. Such an attempt would be equivalent to posing a challenge to the firmly established formulations of quantum theory. The odds of something like this to succeed seem too small to warrant pursuing. Therefore, we argue that the kind of experiment proposed and discussed in the present paper, for which the results are completely predictable by the known properties of quantum mechanics, is the only kind of experiment that can be in principle proposed. The results we described can be considered as mere derivation of the quantum theory.  The reason it is important for us here is that it manifests a perspective important for the current discussion\textthreequartersemdash quantum mechanics may have not left any space for the observer's consciousness to manipulate the experimental results.

This conclusion suggests constraints for understanding the measurement problem and the mental-physical relationship. Firstly, it is necessary to discuss what constitutes a measurement, if we use the collapse of wave function as a defining characteristic of it. Clearly, measurement can be carried out without a macroscopic measuring device. For example, the idler photon, that carries the \textquotedblleft which-path" information, can serve as the measuring device. In similar experiments, atoms with intrinsic states carrying \textquotedblleft which-path" information can also work as measuring devices and hence can cause the interference pattern to disappear \citep{scully_quantum_1982}. Therefore, the suggestion that the measurement is completed when the results are registered in consciousness or when the results are recorded macroscopically (for example, see \citealp{ primas_critical_1997}) does not seem to hold. It appears that neither the conscious registration nor the macroscopic recording is necessary for the collapse of the wave function. Even the interaction with the environment, as suggested by decoherence theory, is not a sufficient ingredient for measurement and collapsing the wave function. Because as long as the \textquotedblleft which-path" is in principle unobtainable, the wave function does not collapse, regardless of the interaction of the system with the environment (e.g., see \citealp{kim_delayed_2000} and other \textquotedblleft quantum eraser" experiments). One alternative is to conceptualize the quantum mechanics as being based on a structure of information \citep{ zeilinger_foundational_1999, brukner_youngs_2002, brukner_quantum_2005}.

Secondly, our argument about the existence of collapse without conscious registration of corresponding results casts strong doubt on those interpretations of quantum mechanics that place the observer's mind in a special position (e.g., many-minds interpretation, \citealp{ lockwood_many_1996}). In such interpretations, the wave function is assumed to be the only and complete description of physical reality. There is no objective \textquotedblleft collapse" occurring outside the mind of the observer. Hence, according to these interpretations, the effect of a measurement is \textquotedblleft to create an entanglement between the state of the system being measured, the measuring apparatus, and the mind of the observer" \citep{ lockwood_many_1996}. Hence, the single state revealed after the measurement is only perceived by the mind and does not reflect any physical event outside this particular mind. However, as we argued, empirical evidence suggests that collapse occurs without the involvement of the mind. This renders the no-collapse-outside-the-mind interpretations untenable. Importantly, the implicit assumption in these interpretations that the mind has a special property and can, through collapse, perceive a single state has led to a symmetrical completion of the relations by proposing that the collapse of the wave function\textthreequartersemdash within the brain\textthreequartersemdash is responsible for the emergence of consciousness (e.g., see \citealp{mensky_reality_2007}). The present analysis provides strong reasons for refuting the underlying arguments: If the former need to be rejected by empirical evidence, the latter loses its foundations.

Thirdly, if consciousness does not play a special role in the measuring procedure, the role of the observer in quantum mechanics would be much less unique or mysterious. The observer would play a role no more special than that in the classical theory, for example, in Einstein's special theory of relativity \citep{shimony_role_1963} or in Darwin's theory of evolution. Some authors suggested that, if consciousness is irrelevant, the role of observer is special in the sense that he/she can choose the quantum reality that will be created. For example, the experimenter may decide whether to realize the interference pattern or not by deciding whether to make the \textquotedblleft which-path" information available or not \citep{brukner_youngs_2002, brukner_quantum_2005}. However, to make such choices special in comparison to other choices made by a mechanically deterministic systems (e.g., a robot) or random number generators (e.g., by playing a quantum dice), one needs to assume that the human observer makes decisions in a qualitatively different way, perhaps through \textquotedblleft free will". Neither theoretical analyses nor the empirical data support the idea that humans make decisions free of the physical processes or of the influences from the environment \citep{ wegner_minds_2003,baum_what_2004, haggard_conscious_2005, haggard_human_2008}. Therefore, we do not see how the fact that the experimenter has a choice could endow him or her with any more special role in the quantum than in the classical theory. Moreover, the conclusion that the observer plays no more a special role in the quantum than in the classical mechanics would hold even if we assumed the existence of \textquotedblleft free will". This is simply because we would then \textquotedblleft create" physical reality routinely, outside the physics experiments, though each individual's actions resulting from our daily interactions with the (mostly non-quantum) world. Therefore, \textquotedblleft free will" cannot save the consciousness hypothesis for the explanation of the measurement problem, nor can it put the human observer at a more special place in quantum theory than it has been assigned in the classical theory.

Finally, it is helpful to note that the current analysis is aimed to clarify a specific relationship between the mind and the physical world, namely the hypothetical necessity of conscious registration of a measurement result to collapse the wave function. We do not try to draw any general conclusions about the relation between physical reality and phenomenal representations. One may argue that, even if we show that the conscious
registration of \textquotedblleft which-path" information is not necessary to collapse the wave function, we cannot exclude the possibility that the consciousness remains nevertheless responsible for the happening of the physical events. For instance, one possibility is that the single-peaked distribution at $D_0$ would never occur without a conscious perception of this distribution. To address this question, it is helpful to clarify the present assumption about physical reality. We used the criteria suggested by Einstein, Podolsky and Rosen \citeyearpar{einstein_can_1935}, stating that if one can predict the physical quantity with certainty and without disturbing the system, one can consider this quantity as a physical reality. In our case, we can consider the distribution at $D_0$ as the physical quantity of interest and we demonstrated that if the \textquotedblleft which-path" information is obtainable, the distribution is always single-peaked (i.e., without interference patterns). That is, in this particular case, we can predict the physical quantity with certainty and without disturbing it. Therefore, we consider the distribution in $D_0$ as physically real. With this clarified, we can infer that the question about the possibility that conscious perception of $D_0$ distribution causes the collapse of all the wave functions simultaneously and creates hence the distribution itself, is an equivalent of asking whether the conscious perception creates the reality of the world in general. According to this latter idea, the universe would not exist if all its conscious creatures close their
eyes and shut their ears. This is a non-trivial question\textthreequartersemdash known as solipsism\textthreequartersemdash that has challenged human intellect for a long time. It is important to point out that this assertion is beyond the scope of scientific enquiry as it is not empirically testable. In other words, there is no conceivable experimental setup by which this statement could be falsified.
Solely for that reason this assertion does not constitute a scientific statement \citep{ popper_logic_1963}.

In conclusion, the available evidence does not indicate that the observer's explicit phenomenal representation about the outcome of a measurement plays a role in collapsing the wave function. We also suggest that the observer does not serve a more fundamental function in quantum mechanics than that in the classical theory. Thus, the idea that by mere observation the experimenter creates physical reality is not viable. This supports Wigner's opinion in his later years and promises to fulfill his hopes\textthreequartersemdash that we \textquotedblleft will not embrace solipsism" and \textquotedblleft will let us admit that the world really exists" (cited from \citealp{ primas_critical_1997}). Perhaps equally importantly, we can add our own hope that the rejection of the role of consciousness in quantum mechanics will also lead us to re-evaluate the proposals that quantum mechanics is vital for explaining the consciousness. Having these two deep mysteries disentangled one from the other might be an important step forward towards understanding better either of them.

\section*{Acknowledgement}
\label{sec:Acknowledgement}
We thank Hrvoje Nikoli\'{c}, Thomas Metzinger, Markus Arndt, Anton Zeilinger, Rajarshi Roy and Nick Herbert for helpful comments.

\bibliographystyle{elsarticle-harv}
\bibliography{Yu_and_nikolic_v2}

\begin{thebibliography}{46}
\expandafter\ifx\csname natexlab\endcsname\relax\def\natexlab#1{#1}\fi
\expandafter\ifx\csname url\endcsname\relax
  \def\url#1{\texttt{#1}}\fi
\expandafter\ifx\csname urlprefix\endcsname\relax\def\urlprefix{URL }\fi

\bibitem[{Baum(2004)}]{baum_what_2004}
Baum, E.~B., 2004. What is thought? The {MIT} Press.

\bibitem[{Bohm(1952)}]{bohm_suggested_1952}
Bohm, D., 1952. A suggested interpretation of the quantum theory in terms of
  {"Hidden"} variables. {I} and {II}. Physical Review 85~(2), 166--193.

\bibitem[{Brukner and Zeilinger(2002)}]{brukner_youngs_2002}
Brukner, C., Zeilinger, A., 2002. Young's experiment and the finiteness of
  information. Philosophical Transactions of the Royal Society A: Mathematical,
  Physical and Engineering Sciences 360~(1794), 1061--1069.

\bibitem[{Brukner and Zeilinger(2005)}]{brukner_quantum_2005}
Brukner, C., Zeilinger, A., 2005. Quantum physics as a science of information.
  In: Quo Vadis Quantum Mechanics? Springer, pp. 47--61.

\bibitem[{Brunner et~al.(2008)Brunner, Branciard, and
  Gisin}]{brunner_possible_2008}
Brunner, N., Branciard, C., Gisin, N., 2008. Possible entanglement detection
  with the naked eye. Physical Review A 78, 052110.

\bibitem[{Chalmers(1997)}]{chalmers_conscious_1997}
Chalmers, D.~J., 1997. The Conscious Mind: In Search of a Fundamental Theory,
  1st Edition. Oxford University Press, {USA}.

\bibitem[{Cramer(1986)}]{cramer_transactional_1986}
Cramer, J.~G., 1986. The transactional interpretation of quantum mechanics.
  Reviews of Modern Physics 58~(3), 647--687.

\bibitem[{D\"{u}rr et~al.(1998)D\"{u}rr, Nonn, and Rempe}]{durr_origin_1998}
D\"{u}rr, S., Nonn, T., Rempe, G., 1998. Origin of quantum-mechanical
  complementarity probed by a \textquoteleft which-way\textquoteright
  experiment in an atom interferometer. Nature 395~(6697), 33--37.

\bibitem[{Eichmann et~al.(1993)Eichmann, Bergquist, Bollinger, Gilligan, Itano,
  Wineland, and Raizen}]{eichmann_youngs_1993}
Eichmann, U., Bergquist, J.~C., Bollinger, J.~J., Gilligan, J.~M., Itano,
  W.~M., Wineland, D.~J., Raizen, M.~G., 1993. Young's interference experiment
  with light scattered from two atoms. Physical Review Letters 70~(16),
  2359--2362.

\bibitem[{Einstein et~al.(1935)Einstein, Podolsky, and
  Rosen}]{einstein_can_1935}
Einstein, A., Podolsky, B., Rosen, N., 1935. Can {Quantum-Mechanical}
  description of physical reality be considered complete? Physical Review
  47~(10), 777--780.

\bibitem[{Esfeld(1999)}]{esfeld_wigners_1999}
Esfeld, M., 1999. Wigner's view of physical reality. Studies In History and
  Philosophy of Science Part B: Studies In History and Philosophy of Modern
  Physics 30, 145--154.

\bibitem[{French(2002)}]{french_phenomenological_2002}
French, S., 2002. A phenomenological solution to the measurement problem?
  {H}usserl and the foundations of quantum mechanics. Studies In History and
  Philosophy of Science Part B: Studies In History and Philosophy of Modern
  Physics 33~(3), 467--491.

\bibitem[{Haggard(2005)}]{haggard_conscious_2005}
Haggard, P., 2005. Conscious intention and motor cognition. Trends in Cognitive
  Sciences 9~(6), 290--295.

\bibitem[{Haggard(2008)}]{haggard_human_2008}
Haggard, P., 2008. Human volition: towards a neuroscience of will. Nature
  Reviews Neuroscience 9~(12), 934--946.

\bibitem[{Kahneman et~al.(1967)Kahneman, Beatty, and
  Pollack}]{kahneman_perceptual_1967}
Kahneman, D., Beatty, J., Pollack, I., 1967. Perceptual deficit during a mental
  task. Science 157~(3785), 218--219.

\bibitem[{Kim et~al.(2000)Kim, Yu, Kulik, Shih, and Scully}]{kim_delayed_2000}
Kim, Y., Yu, R., Kulik, S.~P., Shih, Y., Scully, M.~O., 2000. Delayed
  {"Choice"} quantum eraser. Physical Review Letters 84~(1), 1--5.

\bibitem[{Koch and Hepp(2006)}]{koch_quantum_2006}
Koch, C., Hepp, K., 2006. Quantum mechanics in the brain. Nature 440~(7084),
  611--611.

\bibitem[{Lachter et~al.(2000)Lachter, Durgin, and
  Washington}]{lachter_disappearing_2000}
Lachter, J., Durgin, F., Washington, T., 2000. Disappearing percepts: Evidence
  for retention failure in metacontrast masking. Visual Cognition 7~(1),
  269--279.

\bibitem[{Lockwood(1996)}]{lockwood_many_1996}
Lockwood, M., 1996. {‘Many} minds’ interpretations of quantum mechanics.
  The British Journal for the Philosophy of Science 47~(2), 159 --188.

\bibitem[{London and Bauer(1939)}]{london_la_1939}
London, F., Bauer, E., 1939. La th\'{e}orie de l'observation en m\'{e}canique
  quantique {(Hermann,} Paris). {English} translation in Quantum Theory and
  Measurement, edited by J. A. Wheeler and W. H. Zurek {(Princeton} University,
  Princeton, New Jersey, 1983), pp. 217--259.

\bibitem[{Mandel(1999)}]{mandel_quantum_1999}
Mandel, L., 1999. Quantum effects in one-photon and two-photon interference.
  Reviews of Modern Physics 71~(2), S274--S282.

\bibitem[{Margenau and Wigner(1962)}]{margenau_commentsprofessor_1962}
Margenau, H., Wigner, E.~P., 1962. Comments on professor {Putnam's Comments}.
  Philosophy of Science 29~(3), 292.

\bibitem[{Menskii(2000)}]{menskii_quantum_2000}
Menskii, M.~B., 2000. Quantum mechanics: new experiments, new applications, and
  new formulations of old questions. {Physics-Uspekhi} 43~(6), 585--600.

\bibitem[{Mensky(2007)}]{mensky_reality_2007}
Mensky, M.~B., 2007. Reality in quantum mechanics, extended everett concept,
  and consciousness. Optics and Spectroscopy 103~(3), 461--467.

\bibitem[{Nauenberg(2007)}]{nauenberg_critique_2007}
Nauenberg, M., 2007. Critique of {“Quantum} enigma: Physics encounters
  consciousness”. Foundations of Physics 37~(11), 1612--1627.

\bibitem[{Penrose(2007)}]{penrose_road_2007}
Penrose, R., 2007. The Road to Reality: A Complete Guide to the Laws of the
  Universe. Vintage.

\bibitem[{Popper(1963)}]{popper_logic_1963}
Popper, K., 1963. The Logic of Scientific Discovery. Routledge.

\bibitem[{Primas and Esfeld(1997)}]{primas_critical_1997}
Primas, H., Esfeld, M., 1997. A critical review of {W}igner's work on the
  conceptual foundations of quantum theory. PhilSci Archive.
  http://philsci-archive.pitt.edu/archive/00001574/.

\bibitem[{Putnam(1961)}]{putnam_commentspaper_1961}
Putnam, H., 1961. Comments on the paper of {David Sharp}. Philosophy of Science
  28~(3), 234.

\bibitem[{Putnam(1964)}]{putnam_commentscommentscomments:reply_1964}
Putnam, H., 1964. Comments on {Comments on Comments: a reply to Margenau and
  Wigner}. Philosophy of Science 31~(1), 1.

\bibitem[{Raymond et~al.(1992)Raymond, Shapiro, and
  Arnell}]{raymond_temporary_1992}
Raymond, J.~E., Shapiro, K.~L., Arnell, K.~M., 1992. Temporary suppression of
  visual processing in an {RSVP} task: an attentional blink? Journal of
  Experimental Psychology. Human Perception and Performance 18~(3), 849--860.

\bibitem[{Rensink et~al.(1997)Rensink, {O'Regan}, and Clark}]{rensink_to_1997}
Rensink, R.~A., {O'Regan}, J.~K., Clark, J.~J., 1997. To see or not to see: The
  need for attention to perceive changes in scenes. Psychological Science
  8~(5), 368--373.

\bibitem[{Rosenblum and Kuttner(2008)}]{rosenblum_quantum_2008}
Rosenblum, B., Kuttner, F., 2008. Quantum Enigma: Physics Encounters
  Consciousness. Oxford University Press, {USA}.

\bibitem[{Schr\"{o}dinger(1926)}]{Schrodinger_undulatory_1926}
Schr\"{o}dinger, E., 1926. An undulatory theory of the mechanics of atoms and
  molecules. Physical Review 28~(6), 1049--1070.

\bibitem[{Scully and Dr\"{u}hl(1982)}]{scully_quantum_1982}
Scully, M.~O., Dr\"{u}hl, K., 1982. Quantum eraser: A proposed photon
  correlation experiment concerning observation and \textquotedblleft delayed
  choice" in quantum mechanics. Physical Review A 25~(4), 2208--2213.

\bibitem[{Shimony(1963)}]{shimony_role_1963}
Shimony, A., 1963. Role of the observer in quantum theory. American Journal of
  Physics 31~(10), 755--773.

\bibitem[{Silvanto et~al.(2005)Silvanto, Cowey, Lavie, and
  Walsh}]{silvanto_striate_2005}
Silvanto, J., Cowey, A., Lavie, N., Walsh, V., 2005. Striate cortex {(V1)}
  activity gates awareness of motion. Nature Neuroscience 8~(2), 143--144.

\bibitem[{Stapp(2007)}]{stapp_mindful_2007}
Stapp, H.~P., 2007. Mindful Universe: Quantum Mechanics and the Participating
  Observer. Springer.

\bibitem[{Thaheld(2005)}]{thaheld_does_2005}
Thaheld, F.~H., 2005. Does consciousness really collapse the wave function?:: A
  possible objective biophysical resolution of the measurement problem.
  {BioSystems} 81~(2), 113–124.

\bibitem[{Thaheld(2008)}]{thaheld_modified_2008}
Thaheld, F.~H., 2008. A modified approach to the measurement problem: objective
  reduction in the retinal molecule prior to conformational change. Bio Systems
  92~(2), 114--116.

\bibitem[{Thaheld(2009)}]{thaheld_mechanical_2009}
Thaheld, F.~H., 2009. A mechanical engineer's approach to the measurement
  problem: Is a picture worth a thousand words? arxiv:0909.1526.

\bibitem[{Treisman and Gelade(1980)}]{treisman_feature-integration_1980}
Treisman, A.~M., Gelade, G., 1980. A feature-integration theory of attention.
  Cognitive Psychology 12~(1), 97--136.

\bibitem[{von Neumann(1932)}]{von_neumann_mathematical_1932}
von Neumann, J., 1932. Mathematical Foundations of Quantum Mechanics. 1996
  edition, Beyer, R. T., trans., Princeton Univ. Press.

\bibitem[{Wegner(2003)}]{wegner_minds_2003}
Wegner, D., 2003. The mind's best trick: how we experience conscious will.
  Trends in Cognitive Sciences 7~(2), 65--69.

\bibitem[{Zeilinger(1999{\natexlab{a}})}]{zeilinger_experiment_1999}
Zeilinger, A., 1999{\natexlab{a}}. Experiment and the foundations of quantum
  physics. Reviews of Modern Physics 71~(2), S288--S297.

\bibitem[{Zeilinger(1999{\natexlab{b}})}]{zeilinger_foundational_1999}
Zeilinger, A., 1999{\natexlab{b}}. A foundational principle for quantum
  mechanics. Foundations of Physics 29~(4), 631--643.

\end{thebibliography}

\end{document}